\documentclass[aps,prd,preprint,showpacs,superscriptaddress]{revtex4}
\usepackage{graphics}
\usepackage{graphicx}
\usepackage{pstricks}
\usepackage{epsfig}

\usepackage{multirow}
\usepackage[psamsfonts]{amssymb} 
\usepackage{longtable}

\usepackage{amsmath}
\usepackage{comment}
\usepackage{placeins}
\usepackage{cleveref}

\begin{document}
\newcommand{\du}{$2+1$}

\title{Systematic study of the $SU(3)_c\otimes SU(3)_L\otimes U(1)_X$ local gauge 
symmetry}

\author{Richard H. Benavides} 
\affiliation{Instituto Tecnológico Metropolitano, ITM, 
Facultad de Ciencias Exactas y Aplicadas, calle 73\#76A-354, vía el volador, 
Medellín, Colombia.}
\author{Yithsbey Giraldo}
\affiliation{Departamento de F\'isica, Universidad de Nari\~no, A.A. 1175, San Juan de
Pasto, Colombia.}
\author{Luis Mu\~noz} 
\affiliation{Instituto Tecnológico Metropolitano, ITM, 
Facultad de Ciencias Exactas y Aplicadas, calle 73\#76A-354, vía el volador, 
Medellín, Colombia.}
\author{William A. Ponce}
\affiliation{Instituto de F\'\i sica, Universidad de Antioquia,
A.A. 1226, Medell\'\i n, Colombia.}

\author{Eduardo Rojas}
\affiliation{Departamento de F\'isica, Universidad de Nari\~no, A.A. 1175, San Juan de
Pasto, Colombia.}

\date{\today}

\begin{abstract}
We review in a systematic way how anomaly free $SU(3)_c\otimes SU(3)_L\otimes U(1)_x$ 
models without exotic electric charges can be constructed, using as basis closed sets of 
fermions which includes each one the particles and antiparticles of all the electrically 
charged fields. Our analysis reproduce not only the known models in the literature, 
but also shows the existence of several more independent models for one and three 
families not considered so far. A phenomenological analysis  of the new models is done, where 
the lowest limits at a 95 \% CL on the  gauge boson masses are presented.
\end{abstract}



\maketitle

\section{\label{sec:sec1}Introduction}
The impressive success of the Standard Model (SM) based on the local gauge group 
$SU(3)_c\otimes SU(2)_L\otimes U(1)_Y$ with the color sector $SU(3)_c$ confined 
and the flavor sector $SU(2)_L\otimes U(1)_Y$ hidden and broken spontaneously by 
the minimal Higgs mechanism~\cite{Donoghue:1992dd}, has not been able enough to provide 
explanation for several fundamental issues, among them: the 
hierarchical masses and the mixing angles for both, the quark and  the lepton sectors~\cite{CarcamoHernandez:2015hbk,Csaki:2015xpj,Novichkov:2021evw,Rodejohann:2019izm,Giraldo:2015cpp}, 
charge quantization~\cite{deSousaPires:1998jc,Dong:2005ebq,RomeroAbad:2020uvo,Pisano:1996ht,Doff:1998we,deSousaPires:1999ca,Dong:2005ebq}, the strong CP violation~\cite{Pal:1994ba,Neves:2021usp,Dias:2003zt,Dias:2003iq}, the small neutrino masses and their oscillations~\cite{Boucenna:2014dia,Binh:2021yfy}, 
and last but not least, the presence 
of dark matter and dark energy in the universe~\cite{Kelso:2014qka,Mizukoshi:2010ky,Ruiz-Alvarez:2012nvg,Profumo:2013sca,Filippi:2005mt}. Because of this, many physicists 
believe that the SM does not stand for the final theory, representing only an 
effective model originated from a more fundamental one.

Minimal extensions of the SM arise either by adding new fields, or by enlarging the 
local gauge group (adding a right handed neutrino field constitute its simples 
extension, something that ameliorate, but not solve some of the problems mentioned above).

Simple extensions of the local gauge group consider an electroweak sector with an 
extra abelian symmetry $SU(2)_L\otimes U(1)_x\otimes U(1)_z$~\cite{Ponce:1987wb,Benavides:2018rgh}, or either 
the so called left right symmetric model $SU(2)_L\otimes SU(2)_R\otimes U(1)_{(B-L)}$ 
\cite{Mohapatra:1974gc,Mohapatra:1979ia,Mohapatra:1980qe}, and also 
$SU(3)_L\otimes U(1)_X$, being the last one we are going to consider 
in this study~\cite{Valle:1983dk,Pisano:1992bxx,Frampton:1992wt, Ponce:2001jn,Ponce:2002sg}.

\section{\label{sec:sec2}3-3-1 Models}
In what follows we assume that the electroweak gauge group is  
$SU(3)_c\otimes SU(3)_L\otimes U(1)_X$ (3-3-1 for short) 
in which the electroweak sector of the standard model 
$SU(2)_L\otimes U(1)_Y$ is extended to $SU(3)_L\otimes U(1)_X$. 
We also assume that, as in the SM, the color group $SU(3)_c$ is vector-like (free 
of anomalies) and that the left-handed quarks (color triplets) and left-handed 
leptons (color singlets) transform only under the two fundamental representations 
of $SU(3)_L$ (the 3 and $3^*$). 

Two classes of models will show up: universal one family models where the anomalies 
cancel in each family as in the SM, and family models where the anomalies cancel 
by an interplay between the several families.

For the 3-3-1 models, the most general electric charge operator in the extended 
electroweak sector is 
\begin{equation}\label{qem}
Q=a \lambda_3+ \frac{1}{\sqrt{3}}b\lambda_8 +XI_3,
\end{equation}
where $\lambda_\alpha,\;\alpha=1,2,\dots ,8$ are the Gell-Mann matrices for $SU(3)_L$ 
normalized as Tr$(\lambda_\alpha\lambda_\beta)=2\delta_{\alpha\beta}$ and $I_3=Dg(1,1,1)$ 
is the diagonal $3\times 3$ unit matrix.  If one assumes $a=1/2$, the isospin $SU(2)_L$
of the SM is entirely embedded in $SU(3)_L$; $b$ is a free parameter which fixes the 
model and the $X$ values are obtained by anomaly cancellation. The 8 gauge
fields $A_\mu^\alpha$ of $SU(3)_L$ may be written as~\cite{Ponce:2001jn,Ponce:2002sg}
\begin{equation}\label{gfi}
\sum_\alpha\lambda_\alpha A^\alpha_\mu=\sqrt{2}\left(
\begin{array}{ccc}
D^0_{1\mu} & W^+_\mu & K_\mu^{(b+1/2)} \\
W^-_\mu & D^0_{2\mu} & K_\mu^{(b-1/2)} \\
K_\mu^{-(b+1/2)} & K_\mu^{-(b-1/2)} & D^0_{3\mu} \\
\end{array}\right),
\end{equation}
where $D^0_{1\mu}=A_\mu^3/\sqrt{2}+A_\mu^8/\sqrt{6},\; 
D^0_{2\mu}=-A_\mu^3/\sqrt{2}+A_\mu^8/\sqrt{6},$ and 
$D^0_{3\mu}= -2A_\mu^8/\sqrt{6}.$  The upper indices on the gauge bosons in Eq.~\eqref{gfi} stand for 
the electric charge of the particles, some of them being functions of the $b$ parameter~\cite{Byakti:2020ipa}.

\section{\label{sec:sec3}The Minimal Model}
In Ref.~\cite{Valle:1983dk,Frampton:1992wt} it has been shown that, for $b=3/2$, the following fermion 
structure is free of all the gauge anomalies: 
$\psi_{lL}^T= (\nu_l^0,l^-,l^+)_L\sim (1,3,0),\; Q_{iL}^T=(d_i,u_i,X_i)_L\sim (3,3^*,-1/3),\; 
Q_{3L}^T=(u_3,d_3,Y)\sim (3,3,2/3)$, where $l=e,\mu,\tau$ is a family lepton 
index, $i=1,2$ for the first two quark families, and the numbers after the 
similarity sign means 3-3-1 representations. The right handed fields are 
$u_{aL}^c\sim (3^*,1,-2/3),\; d_{aL}^c\sim (3^*,1,1/3),\; X_{iL}^c\sim(3^*,1,4/3)$ 
and $Y_L^c\sim (3^*,1,-5/3)$, where $a=1,2,3$ is the quark family index, and there 
are two exotic quarks with electric charge $-4/3\; (X_i)$ and other 
with electric charge 5/3 $(Y)$. This version is called minimal in the literature, 
because it does not make use of exotic leptons,
including possible right-handed neutrinos.

\section{\label{sec:sec4}
3-3-1 Models Without Exotic Electric Charges}
If one wishes to avoid exotic electric charges in the fermion and boson sectors as 
the ones present in the minimal~(3-3-1) model, one must choose $b=1/2$ in Eq. 
(\ref{qem}) as shown in Ref~\cite{Ponce:2001jn,Ponce:2002sg}.

To begin with our systematic analysis, let us start with closed fermion structures 
consisting of only one left handed $SU(3)_L$ triplet and right handed singlets, where 
for ``closed'' we mean structures containing the antiparticles of all the 
electric charged particles. Following the notation in Ref.~\cite{Ponce:2001jn,Ponce:2002sg}, there are only 
six of such structures containing at least one of the fermion fields in one family of the SM, or in its minimal extension with right-handed neutrino fields:

there are only six such structures that contain at least one of the fermion fields in a family of SMthere are only six such structures that contain at least one of the fermion fields in a family of SM

\begin{itemize}
\item $S_1=[(\nu^0_e,e^-,E_1^-)\oplus e^+\oplus E_1^+]_L$ 
with quantum numbers $(1,3,-2/3);(1,1,1)$ and $(1,1,1)$ respectively.
\item $S_2=[(e^-,\nu_e^0,N_1^0)\oplus e^+]_L$ 
with quantum numbers $(1,3^*,-1/3)$ and $(1,1,1)$ respectively.
\item $S_3=[(d,u,U)\oplus u^c\oplus d^c\oplus U^c]_L$ with quantum numbers 
$(3,3^*,1/3)\;$; $(3^*,1,-2/3)\;$; $(3^*,1,1/3)$ and $(3^*,1,-2/3)$ respectively.
\item $S_4=[(u,d,D)\oplus u^c\oplus d^c\oplus D^c]_L$ with quantum numbers 
$(3,3,0)\;$; $(3^*,1,-2/3)\;$; $(3^*,1,1/3)$ and $(3^*,1,1/3)$ respectively.
\item $S_5=[(N_2^0, E_2^+,e^+)\oplus E^-_2\oplus e^-]_L$ with quantum numbers 
$(1,3^*,2/3)\;\; (1,1,-1)$, and $(1,1,-1)$ respectively.
\item $S_6=[(E^+_3,N_3^0,N_4^0)\oplus E_3^-]_L$ with quantum numbers $(1,3,1/3)$ 
and $(1,1,-1)$ respectively,
\end{itemize}
where for phenomenological reasons we allow for the precense of several exotic 
leptons (charged and neutral), but only one exotic quark of each type~(down-type or up-type).
In the former sets, $N^0_1$ and $N_4^0$ can play the role of the right handed neutrino 
field $\nu_e^{0c}$ in an SO(10) basis.

Notice that the value $b=1/2$ in Eq.~\eqref{qem} implies that the electric charge of the last two components of a 3 or a $3^*$ of $SU(3)_L$ are of the same value.
At this point, our approach is different to the one presented in 
Ref.~\cite{Ponce:2001jn,Ponce:2002sg}, the difference being that only one fundamental representation of $SU(3)_L$~(triplet or anti-triplet) is used 
in each set, instead of the composite ones present in the original reference~\cite{Valle:1983dk,Pisano:1992bxx,Frampton:1992wt} 
(such composite lepton structures will appear anon in our systematic analysis).

\begin{table}
\scalebox{1.1}{
\begin{tabular}{||l|cccccc||}\hline\hline
Anomalies & $S_1$ & $S_2$ & $S_3$ & $S_4$ & $S_5$ & $S_6$ \\ \hline
$[SU(3)_C]^2U(1)_X$ & 0 & 0 & 0 & 0 & 0 & 0 \\
$[SU(3)_L]^2U(1)_X$ & $-2/3$  & $-1/3$ & 1 & 0& 2/3 & 1/3\\
$[Grav]^2U(1)_X$ & 0 & 0 & 0 & 0 & 0 & 0 \\
$[U(1)_X]^3$ & 10/9 & 8/9 & $-4/3$ & $-2/3$& -10/9& -8/9 \\
$[SU(3)_L]^3$ & 1 & $-1$ & $-3$ & 3 & $-1$ & 1\\
\hline\hline
\end{tabular}} 
\caption{Anomalies for some 3-3-1 fermion fields structures}
\label{tabl1}
\end{table}
\noindent
The several gauge anomalies calculated for these six sets are shown in Table~\ref{tabl1}; 
where notice that the anomaly values for $S_1,\; S_2,\; S_3$ and $S_4$ coincide with 
the ones presented in Ref~\cite{Ponce:2001jn,Ponce:2002sg}, being the values for $S_5$  and $S_6$ new results.

Now, if we want to consider only one family of quarks, either the sets $S_3$  or $S_4$ 
are enough, but for 3 quark families, one of the following combinations must be used:
$3S_3,\; 3S_4, (2S_3+S_4)$ and $(S_3+2S_4)$, where the first two ones are 
associated with universal models in the quark sector.

Right from Table 1 it is simple to read the following sets free of 
anomalies: 
\begin{itemize}
\item {\bf Model I:}\hspace{1cm}
$2S_2+S_4+S_5$,
\item{\bf Model J:}\hspace{1cm}
$2S_1+S_3+S_6$,
\item{\bf Model A:}\hspace{1.2cm}
$3S_2+S_3+2S_4$,
\item{\bf Model B:}\hspace{1.2cm}
$3S_1+2S_3+S_4$,
\end{itemize}
where the structures {\bf I} and {\bf J}~\cite{acad} contain only one family of quarks,  
and {\bf A} and {\bf B} are three family quark models~(here we are following the notation in Ref.~\cite{acad}). But, can we view {\bf I}  
and {\bf J} as one family (``universal'') anomaly free models? the answer is 
yes if we allow models with exotic electrons and new electric neutral particles. 
As a matter of fact, we can writte the particle content for Model {\bf I} 
as~\cite{acad}:
\begin{eqnarray*}
[(e^-,\nu^0_e,N^0_1)&\oplus& e^+\oplus 
(E^-_1,N^0_2,N^0_3)\oplus E^+_1\oplus \\
(N^0_4,E^+_2,E^+_3)&\oplus& E_2^-\oplus E^-_3\oplus (u,d,D)
\oplus u^c\oplus d^c\oplus D^c]_L.
\end{eqnarray*}
Ugly as it may be, due to the precense of several exotic leptons, some with the same quantun 
numbers of the ordinary ones, we may said that this model is not yet excluded from 
present phenomenology.

In a similar way, the particle content of the structure {\bf J} can be written 
as~\cite{acad}:
\begin{eqnarray*}
[(\nu^0_e,e^-,E_1^-)&\oplus& e^+\oplus E^+_1\oplus 
(N_1,E^-_2,E^-_3)\oplus E^+_2\oplus E^+_3\oplus \\
(E^+_4,N_2^0,N_3^0)&\oplus& E_4^-\oplus (d,u,U)
\oplus u^c\oplus d^c\oplus U^c]_L.
\end{eqnarray*}

The other two structures {\bf A} and {\bf B} correspond to two well known non 
universal models already present in the literature; {\bf A} being named as 
a ``3-3-1 model with right-handed neutrinos''~\cite{Montero:1992jk,Foot:1994ym,Benavides:2009cn} and {\bf B} named as 
a ``3-3-1 model with exotic charged leptons''~\cite{Ozer:1995xi,Ponce:2006au,Salazar:2007ym}.

An unrealistic two family anomaly free structure is for example $S_1+S_2+S_3+S_4$ 
(unrealistic because there is strong evidence for at least three families 
in nature~\cite{Donoghue:1992dd}).

The next  strategy is to use the lepton sets $S_1,\; S_2,\; S_5$ and $S_6$ to 
build non vector-like new sets of leptons~(vector-like sets are free of anomalies by definition~\cite{Donoghue:1992dd},
and quark sector anomalies must cancel out with those of the lepton sector). Notice that vector-like sets as 
for example $S_1+S_5$ and $S_2+S_6$ are free of anomalies and not suitable 
for constructing realistic models due to the non zero anomalies in the 
quark sector. 
For the same reason we exclude from our analysis vector like structures as 
$(\nu^0_e,e^-,E^-)\oplus (N^0,e^+,E^+)\sim (1,3,-2/3)\oplus (1,3^*,2/3)$ and 
$(e^-,\nu^0_e,N^0_1)\oplus (E^+,N^0_2,N^0_3)\sim (1,3^*,-1/3)\oplus (1,3,1/3)$.

To be systematic, let us start first with sets of leptons containing 
only two $SU(3)_L$ triplets or anti-triplets:
\begin{itemize}
\item $S_7=[(e^-,\nu^0_e,N_1^0)\oplus (N_2^0,E^+,e^+)\oplus E^-]_L$ 
with quantum numbers $(1,3^*,-1/3)\;$; $(1,3^*,2/3)$ and $(1,1,-1)$ respectively.
\item $S_8=[(\nu^0_e,e^-,E^-)\oplus (E^+,N_1^0,N_2^0)\oplus e^+]_L$ 
with quantum numbers $(1,3,-2/3)$, $(1,3,1/3)$ and $(1,1,1)$ respectively.
\end{itemize}
Notice again $S_7+S_8$ is vector-like and in consequence is free of anomalies 
and unsuitable for our building process.

The next step is to include lepton sets with three $SU(3)_L$ triplets or 
anti-triplets: 
\begin{itemize}
\item $S_9=[(e^-,\nu_e,N_1^0)\oplus (E^-,N_2^0,N_3^0)\oplus (N_4^0, E^+,e^+)]_L$ 
with quantum numbers $(1,3^*,-1/3)$;$(1,3^*,-1/3)$ and $(1,3^*,2/3)$ respectively.
\item $S_{10}=[(\nu_e, e^-,E_1^-)\oplus (E^+_2,N_1^0,N_2^0)\oplus 
e^+ \oplus (N_3^0, E^-_2,E_3^-)\oplus E_1^+\oplus E_3^+]_L$ with quantum numbers 
$(1,3,-2/3);\;\;(1,3,1/3);\; (1,1,1);\; (1,3,-2/3)$; 
$(1,1,1)$, and $(1,1,1)$ respectively.
\item $S_{11}=[(e^-,\nu_e,N_1^0)\oplus (N_2^0,E^+_1,e^+)\oplus 
(N_3^0, E^+_2,E^+_3)\oplus E^-_1\oplus E^-_2\oplus E^-_3]_L$ 
with quantum numbers $(1,3^*,-1/3);\;\;(1,3^*,2/3);\;\; (1,3^*,2/3);\;\;(1,1,-1);\;\;$ 
$(1,1,-1)$, and $(1,1,-1)$ respectively.
\item $S_{12}=[(\nu_e^0, e^-,E_1^-)\oplus (E^+_1,N_1^0,N_2^0)\oplus 
(E^+_2,N^0_3,N^0_4)\oplus e^+\oplus E_2^-]_L$ with quantum numbers 
$(1,3,-2/3)$; $(1,3,1/3);\; (1,3,1/3);\;\; (1,1,1)$, and $(1,1,-1)$; respectively.
\end{itemize}
The anomalies for these new lepton sets are given in Table~\ref{tabl2}.

At this step we can stop combining new sets, and analyze the rich structure we have 
gotten so far with the anomaly values presented in the former two Tables:

\begin{table}
\scalebox{1.1}{
\begin{tabular}{||l|cccccc||}\hline\hline
Anomalies & $S_7$ & $S_8$ & $S_9$ & $S_{10}$ & $S_{11}$ & $S_{12}$ \\ \hline
$[SU(3)_C]^2U(1)_X$& 0 & 0 & 0 & 0 & 0 & 0\\
$[SU(3)_L]^2U(1)_X$ & 1/3 & -1/3& 0 & -1 & 1 & 0\\
$[Grav]^2U(1)_X$ & 0 & 0 & 0 & 0 & 0 & 0\\
$[U(1)_X]^3$  & $-2/9$ & $2/9$& 2/3& 4/3 & -4/3 & -2/3 \\
$[SU(3)_L]^3$ & $-2$ & 2 & $-3$ & 3 & $-3$ & 3 \\
\hline\hline
\end{tabular} }
\caption{Anomalies for the 3-3-1 non Vector-Like lepton fields structures}
\label{tabl2}
\end{table}

A simple computer program allow us to construct Table~\ref{tabl3}, which by the way is 
our main result of the first part of this paper. Let us see: 

\section{Irreducible Anomaly free sets}
Table~\ref{tabl3} lists all the basic and irreducible sets of multiplets of quarks and leptons which are free 
of anomalies (hereafter, Irreducible Anomaly-Free-Set~(IAFS)), classified according to their 
quark content. These sets can be 
combined in several different ways in order to construct anomaly free three family 
models.
In the first column in Table~\ref{tabl3} the index $i$ lists the IAFSs, 
the second column shows various possible irreducible lepton sets free of anomalies. A closer look
to the sets of the second column shows that all of them are vector-like structures, not 
suitable to build simple models, but useful to build complex anomaly free models, as we 
will see ahead.
Column three contains one of the most important results of our analysis. As a
matter of fact, it lists 20 universal models (one family models), where only four of them are 
reported in the literature so far. Let us see:

$Q^I_1=S_4+S_9$ called {\it carbon copy one} or {\bf ``Model G''} in Ref.~\cite{Ponce:2001jn,Ponce:2002sg}. 
This structure can be embedded in the unification group E(6) according to 
Ref.~\cite{Sanchez:2001ua}, with some phenomenology of this structure already presented 
in the same reference.

$Q^I_2=S_3+S_{10}$ called {\it carbon copy two} or {\bf ``Model H''} in Ref.~\cite{Ponce:2001jn,Ponce:2002sg}. 
It can be embedded in the unification group $SU(6)\otimes U(1)$ according to 
Ref.~\cite{Martinez:2001mu}, with some phenomenology of this structure already presented 
in the same reference.

$Q^I_5=2S_1 + S_3+S_6$ and $Q^I_6=2S_2 + S_4+S_5$ were introduced in Re.~\cite{acad} 
where they were named as {\bf ``Model J''} and {\bf ``Model I''}, respectively.

The other 16 structures correspond to totally new, one family (universal) models, 
which will be analyzed in the  second part of this paper.
But there is much more in the third column of Table~\ref{tabl3}: we can construct three family models using 
sets in column three
not only by iterating three times every structure, but also combining them by 
taking two equal and one diferent (producing in this way 380 three family models),
or by combining three different sets (producing in this way 1140 three family models). 
Let us see some of them:\\
$2Q^I_1 +Q^{I}_2=2S_4+2S_9+S_3+S_{10}$ which is denoted as {\bf Model E} in Ref.~\cite{Ponce:2001jn,Ponce:2002sg},
called {\it hybrid model one} in that paper.\\
$Q^I_1 +2Q^{I}_2=S_4+S_9+2S_3+2S_{10}$ which is denoted as {\bf Model F} in Ref.~\cite{Ponce:2001jn,Ponce:2002sg},
called {\it hybrid model two} in that paper.

Column four in Table~\ref{tabl3} shows an amusing result; aparently it is related to unphysical 
structures (only two families of quarks). But the point is that each one of the 
seven entries in this column can be combined with each one of the 20 entries in the 
third column, in order to produce a three family quark structure, for a total of 140 
anomaly free three family models, most of them new ones. Let us see some examples:\\
$Q^I_1 +Q^{II}_1=S_1+S_2+S_3+2S_4+S_9$ which is denoted as {\bf Model C} in Ref.~\cite{Ponce:2001jn,Ponce:2002sg},  
called {\it a model with unique lepton generation one} in such paper.\\
$Q^I_2 +Q^{II}_1=S_1+S_2+2S_3+S_4+S_{10}$ which is denoted as {\bf Model D} in Ref.~\cite{Ponce:2001jn,Ponce:2002sg}, 
called {\it a model with unique lepton generation two} in the same paper.

The last column in Table~\ref{tabl3} shows the simplest and most economical three family 
structures available for 3-3-1 models without exotic electric charges. As a matter 
of fact, $Q^{III}_1$ is the well known 3-3-1 model with right handed 
neutrinos~\cite{Montero:1992jk,Foot:1994ym,Benavides:2009cn}, and $Q^{III}_2$ is the 3-3-1 model with exotic 
electrons~\cite{Ozer:1995xi,Ponce:2006au,Salazar:2007ym}. All of these are well known models and they are summarized in Table~\ref{tabl4}.

At this point we can use the second column in Table~\ref{tabl3}, by adding one or more 
entries in the Table to any anomaly free three family structure constructed 
from the other three columns in the way we have explained it above. In this way we end up with an immeasurable
number of anomaly free fermion structures, constructed out of the 12 fermion structures~(10
lepton and two quark) introduced in Sec.~\ref{sec:sec4}.

Model count summary:
leaving aside the IAFSs of the second column in Table~\ref{tabl3}, we can make a very illuminating analysis to exemplify the  model building process.
Each one of the IAFSs in the third column~($Q^I$) in Table~\ref{tabl3} contains, at least, one family of SM leptons and quarks, so that, if we threefold this structure, one IAFS for each family, it is possible to have a universal model. Following this procedure, we can obtain 20 models. Another possibility is to choose the same IAFSs for two families and a different one for the third family, in this case we have $19\times20=380$ possibilities. Taking a different IAFS for each family we will have $\binom{20}{3}=1140$ different models. Adding these results with the 2 models coming from  the fifth column in Table~\ref{tabl3},  and the $7\times20=140$ combinations between the IAFSs in the columns $Q^{I}$ and $Q^{II}$, we obtain  a total of 1682 models with a minimum content of exotic quarks. It is clear that from the arbitrary union of  IAFSs  it is possible to obtain an infinite number of models, but in general, these models will have too many exotic particles.\\
\begin{table}[h!!!]
\centering
\begin{tabular}{c|cccc}
\hline
\multicolumn{5}{c}{Irreducible anomaly free sets} \\
\hline
&&&\\[-7mm]
$i$&Vector-like lepton sets~($L_i$) & One quark set~($Q^I_i$)&Two quark sets~($Q^{I\! I}_i$)& Three
quark 
sets~($Q^{I\! I\! I}_i$)\\\hline
1&$S_1+S_5$&$S_4+S_9$&$S_1+S_2+S_3+S_4$&$3 S_2+S_3+2 S_4$\\
2&$S_2+S_6$&$S_3+S_{10}$&$2 S_1+S_3+S_4+S_7$&$3 S_1+2 S_3+S_4$\\
3&$S_7+S_8$&$S_2+S_4+S_7$&$2 S_2+S_3+S_4+S_8$&\\
4&$S_{10}+S_{11}$&$S_1+S_3+S_8$ &$3 S_{2}+S_{3}+S_{4}+S_{12}$&\\
5&$S_9+S_{12}$&$2S_1+S_3+S_6$&$3 S_{1}+2 S_{3}+S_{12}$&\\
6& $S_1+S_6+S_7$&$2S_2+S_4+S_5$&$3 S_{2}+2 S_{4}+S_{11}$&\\
7&$S_6+S_8+S_9$&$S_1+S_4+2 S_7$&$3 S_{1}+S_{3}+S_{4}+S_{11}$&\\
8&$S_2+S_5+S_8$&$S_2+S_3+2 S_8$&&\\
9&$S_5+S_7+S_{10}$&$S_1+S_2+S_3+S_{12}$&&\\
10&$S_2+S_7+S_{12}$&$S_1+S_2+S_4+S_{11}$&&\\
11&$S_1+S_8+S_{11}$&$S_4+3 S_7+S_{10}$&&\\
12& $S_1+2S_6+S_9$&$S_3+3 S_8+S_9$&&\\
13&$S_6+2 S_7+S_{10}$&$2 S_{1} + S_{3} + S_{7} + S_{12}$&&\\
14& $S_5+2 S_8+S_9$&$2 S_{1} + S_{4} + S_{7} + S_{11}$&&\\
15&$S_5+S_6+S_9+S_{10}$&$2 S_{2} + S_{3} + S_{8} + S_{12}$&&\\
16&$S_2+2 S_5+S_{10}$&$2 S_{2} + S_{4} + S_{8} + S_{11}$&&\\
17&$S_{1} + 2 S_{7} + S_{12}$&$3 S_{2} + S_{3} + 2 S_{12}$&&\\
18&$S_{1} + S_{2} + S_{11} + S_{12}$&$3 S_{2} + S_{4} + S_{11} + S_{12}$&&\\
19&$S_{2} + 2 S_{8} + S_{11}$&$3 S_{1} + S_{3} + S_{11} + S_{12}$&&\\
20&$2 S_{1} + S_{6} + S_{11}$&$3 S_{1} + S_{4} + 2 S_{11}$&&\\
21&$2 S_{2} + S_{5} + S_{12}$&&&\\
\hline
\end{tabular}
\caption{ IAFSs. 
Any general Anomaly Free-Set~(AFS) containing quarks, must be a combination of IAFSs~(i.e., $L_i$, $Q^I$, $Q^{II}$ and 
$Q^{III}$)  even for more than three families.
For leptons, the second column~(L) is not exhaustive and it was not possible to account for all the possibilities.}
\label{tabl3}
\end{table}

\begin{table}[h]
\centering
\begin{tabular}{ccc}
\hline
\quad\quad Name\quad\quad & \quad\quad Model\quad\quad&\quad\quad AFS\quad\quad\\\hline
Model {\bf A}&$Q^{I\! I\! I}_1$&$3S_2+S_3+2S_4$\\
Model {\bf B}&$Q^{I\! I\! I}_2$&$3S_1+2S_3+S_4$\\
Model {\bf C}&$Q^I_1+Q^{I\! I}_1$&$S_1+S_2+S_3+ 2S_4+S_9$\\
Model {\bf D}&$Q^I_2+Q^{I\! I}_1$&$S_1+S_2+2S_3+ S_4+S_{10}$\\
Model {\bf E}&$Q^I_2+2\:Q^I_1$&$2S_4+ 2S_9+S_3+S_{10}$\\
Model {\bf F}&$2\:Q^I_2+Q^I_1$&$S_4+ S_9+2S_3+2S_{10}$\\
Model {\bf G}&$3\:Q^I_1$&$3(S_4+S_9)$\\
Model {\bf H}&$3\:Q^I_2$&$3(S_3+S_{10})$\\
Model {\bf I}&$3\:Q^I_6$&$3(2S_2+S_4+S_5)$\\
Model {\bf J}&$3\:Q^I_5$&$3(2S_1+S_3+S_6)$\\
\hline
\end{tabular}
\caption{The 3-3-1 models already reported in the literature. Particular embeddings are assumed for the AFSs, which are well known in the literature~\cite{Ponce:2001jn,Ponce:2002sg}.}
\label{tabl4}
\end{table}

\section{Embeddings of the SM fermions}
From the previous section, we have several anomaly-free representations for the 3-3-1 gauge group. 
From now on we will make a distinction between an Anomaly Free-Set~(AFS) and   a particular embedding corresponding to a distinguishable phenomenological model.
In these representations, there are several ways to assign the SM particles within the available multiplets. For the quark sector, this identification is easy since there are only two sets of multiplets of quarks, ${S}_3$ and ${S}_4$, depending on whether the SM quark doublet is within a $SU(3)_L$ triplet or an anti-triplet. For the lepton sector, there are several sets of multiplets $S_i$ by allowing the right-handed charged lepton to be the third component of a $SU(3)_L$ triplet or a singlet. Once the $S_i$ are chosen for quarks and leptons, there are still several choices for the SM particles within the multiplets.  

In order to illustrate the possible embeddings, we will consider the  AFSs associated with the models~{\bf A}, {\bf E}, {\bf G}, $ Q^ I_3 $, {\bf I} and $ Q^ I_7 $ .
For model~{\bf A}, its particle content is indicated in Table~\ref{tabl4}, and there is only one possible identification for the SM particles in the lepton sector~($3S_2^{\ell + e^+} $), in this case we have three identical copies of the set of multiplets $S_2$. In this case the SM left-handed lepton doublet~($\ell$) is embedded in the $SU (3)_L $ triplet of $S_2$ and the right-handed charged lepton~($e^+$) in the singlet.
Additionally, the model contains three exotic neutral particles, $N^0$.
In Tables \ref{tabl5}  and \ref{tabl6} the SM particle content is shown in the superscript of the fermion sets $S_i$, avoiding  any mention of exotic particles. 

Another interesting example is the model~{\bf E} which has a rich content of particles
and therefore its corresponding AFS  has many possible embeddings, as shown in Table~\ref{tabl4} and \ref{tabl5}.  For example, we can choose the following embedding for the SM particles:
\begin{equation*}
\begin{split}
2S_9+S_{10}&=2\left[\underbrace{(e^-,\nu_e^0,N_1^0)}_{\textrm{SM}}\oplus (N_4^0,E^+,\underbrace{e^+}_{\text{SM}})\oplus(E^-,N_2^0,N_3^0)\right]_L\\
&+\left[\underbrace{(\nu_e,e^-,E_1^-)\oplus e^+}_{\textrm{SM}}
\oplus(E^+_2 , N_1^0 , N_2^0)\oplus
(N_3^0,E_2^-,E_3^-)\oplus E_1^+\oplus E_3^+\right]_L.
\end{split}
\end{equation*}
In this embedding, two SM lepton families are put in $2S_9$ and one SM family in $S_{10}$. In  $ S_9 $ the right-handed lepton is in the third component of an anti-triplet. The third family is embedded into
$S_{10}$ where the right-handed charged lepton is a singlet.
The sets, $ S_9 $ and $ S_{10} $, have different particle content and quantum numbers. This particular embedding is usually known as the model~{\bf E}, which is not universal in the lepton sector and it corresponds to the model $E^{1}$ in Table~\ref {tabl5}.

In the two sets of particles, $ 2S_9 $, there are 4 anti-triplets of $SU(3)_L $ with identical quantum numbers, by  embedding the three left-handed lepton doublets of the SM in these anti-triplets and the three SM right-handed leptons into the $ S_ {10} $ singlets, we obtain a new embedding which corresponds to the model $E^5$ in Table~\ref {tabl5},  as follows:

\begin{equation*}
\begin{split}
&\left[\underbrace{(e^-,\nu_e^0,N_1^0)\oplus(E^-,N_2^0,N_3^0)\oplus(e^-,
\nu_e^0,N_1^0)}_{\textrm{
SM}}\oplus(E^-,N_2^0,N_3^0)\oplus 
(N_4^0,E^+,e^+)\oplus 
(N_4^0,E^+,e^+)\right]_L\\
&+
\left[\underbrace{e^+\oplus E_1^+\oplus
E_3^+}_{\textrm{SM}}
\oplus(\nu_e,e^-,E_1^-)\oplus(E^+_2 , N_1^0 , N_2^0)\oplus 
(N_3^0,E_2^-,E_3^-) \right]_L\, .
\end{split}
\end{equation*}
At a phenomenological level, the last embedding, which is equivalent to the model~{\bf A}, exceeds it by the exotic vector-likes:
$(E^-,N_2^0,N_3^0)\oplus (E^+_2 , N_1^0 , N_2^0)$, 
$(N_4^0,E^+,e^+)\oplus (\nu_e,e^-,E_1^-)$ and $(N_4^0,E^+,e^+)\oplus 
(N_3^0,E_2^-,E_3^-)$. 
This result implies that adding vector-like lepton content to an anomaly-free set of fermions could result in a different non-trivial model by taking a different embedding. 
The embeddings for the models~$I^{j}$ and $Q_7^{Ij}$ are given in the respective Tables~\ref{tabl5} and~\ref{tabl6}. These models are always universal in the quark sector, which is very convenient to avoid FCNC.  There are four $I^j$ embeddings, two universal and two non-universal as shown in Table~\ref{tabl5}. The $I^1$ model  has the same SM particle content as model~{\bf A},  with additional exotic leptons.
 There are 20 possible $Q ^ {Ij}_{7} $ embeddings, with only 4 of them universal, as listed in Table~\ref{tabl6}.
Interesting examples of AFSs that do not contain the vector-like structures listed in column~2 in Table~\ref{tabl3} are:  model~{\bf G} and the models~$Q_3 ^{Ij}$. Model {\bf G} is universal and it has only one embedding, while the~$Q_3^{Ij} $ models are universal in the quark sector and for some of its embeddings in the lepton sector.

\section{Collider Constraints}
In general, each one of the possible embeddings has a different phenomenology. For collider constraints only the $Z'$ couplings to the SM particles matter~\cite{Langacker:2008yv,Cao:2016uur,Fonseca:2016tbn}. In the lepton sector, these couplings depend on whether left-handed lepton doublet $\ell$ is in a $SU(3)_L$ triplet or anti-triplet. In the same way, the $Z'$ coupling for the right-handed charged lepton $e^+$ depends on whether  is a  $SU(3)_L$ singlet or it is the third component of a $SU(3)_L$ anti-triplet.  Similar caveats hold for the SM quark doublets $q$ and the right-handed quarks as shown in the Table~\ref{tabla8}. 

We obtain the lower limit on the $Z'$ mass in Table~\ref{tabla8},  from the intersection of the 95\% CL upper limit on the cross-section from searches of high-mass dilepton resonances at the ATLAS experiment~\cite{ATLAS:2019erb}
with the theoretical cross-section reported in~\cite{Erler:2011ud}. 
In the reference~\cite{ATLAS:2019erb} the upper limits on the cross-section go up to 6~TeV, however, we extrapolate up to 7~TeV in order to obtain the restrictions for the simplified models $C_4+\bar{q}$ and $C_4+q$ in Table~\ref{tabla8}.
The cross-section depends on  the $Z^\prime$ charges as they are given the Appendix~\ref{sec:apendice1}.   The  ATLAS data was obtained from proton-proton collisions at a center-of-mass energy of $\sqrt{s}=13$~TeV during Run 2 of the Large Hadron Collider and correspond to an integrated luminosity of 139 fb$^{-1}$.  Further details are shown in references~\cite{Erler:2011ud,Salazar:2015gxa,Benavides:2018fzm}.
We obtain the constraints in Tables~\ref{tabl5} and~\ref{tabl6} from those shown in Table~\ref{tabla8}.  We only report lower limits for embeddings for which it is possible to choose the same $Z'$ charges for the first two families. Under these assumptions, for the models {\bf A}, $C^j$, {\bf E} , {\bf G}, {\bf I}, $Q_3^{I_j}$ and $Q_7^{I_j}$,  in Tables~\ref{tabl5} and~\ref{tabl6}, the left-handed quark doublet $q$ is part of a $SU(3)_L$ triplet. 
For the remaining models, in the mentioned tables, $q$ is part of a $SU(3)_L$ anti-triplet. 
\begin{table}[h!]
\begin{center}
\begin{tabular}{|c| c | c |c|}
\hline
\hline 
Model     & SM Particle  &Short     &    LHC-Lower limit   \\  [-0.2cm]
          & embedding    &Notation  &     in TeV         \\[0.2cm] \hline 
$C_1+{\bar{q}}$ &$\ell\subset 3$,  \ $e^{+}\subset 1$,  \ $q\subset 3^*$  & $\ell$+$e^{+}$+$\bar{q}$              &  5.53 \\ \hline
$C_1+{q}$       &$\ell\subset 3$,  \ $e^{+}\subset 1$,  \ $q\subset 3 $   & $\ell$+$e^{+}$+$q$                    &  5.33  \\ \hline
$C_2+{\bar{q}}$ &$\ell\subset 3^*$,\ $e^{+}\subset 1$,  \ $q\subset 3^*$  & $\bar{\ell}$+$e^{+}$+$\bar{q}$        &  4.98 \\ \hline
$C_2+{q}$       &$\ell\subset 3^*$,\ $e^{+}\subset 1$,  \ $q\subset 3 $   & $\bar{\ell}$+ $e^{+}$+$q$             &  4.87     \\ \hline
$C_3+{\bar{q}}$ &$\ell\subset 3^*$,\ $e^{+}\subset 3^*$,\ $q\subset 3^* $ & $\bar{\ell}$+ $e^{\prime+}$+$\bar{q}$ &  5.75\\ \hline
$C_3+{q}$       &$\ell\subset 3^*$,\ $e^{+}\subset 3^*$,\ $q\subset 3$  & $\bar{\ell}$+ $e^{\prime+}$+$q$       &  5.53    \\ \hline 
$C_4+{\bar{q}}$ &$\ell\subset 3$,  \ $e^{+}\subset 3^*$,\ $q\subset 3^*$  & $\ell$+ $e^{\prime+}$+$\bar{q}$       &  7.00     \\ \hline
$C_4+{q}$       &$\ell\subset 3$,  \ $e^{+}\subset 3^*$,\ $q\subset 3$    & $\ell$+ $e^{\prime+}$+$q$             &  6.52       \\ \hline
\end{tabular}
\caption{
Simplified 3-3-1 models. The collider constraints depend on whether the SM left-handed doublets $\ell$ or $q$ are contained in a $SU(3)_L$  triplet or anti-triplet. In short notation $q$, $\bar{q}$  or $\ell$, $\bar{\ell}$.
The notation $e^{+\prime}$ is used  if the right-handed lepton is the third component of an anti-triplet,  otherwise $e^{+}$. 
In the first column, $ C_1 $ corresponds to the embedding
$\ell\subset 3$,  \ $e^{+}\subset 1$,
in a similar way are defined the remaining $C_i$.
The $Z'$ charges for the $C_i$, $q$ and $\bar{q}$ are given in~\Cref{tab:c1,tab:c2,tab:c3,tab:c4,tab:qbar,tab:q}, respectively.
In the last column the 95\% CL lower limits on the $Z'$ mass, coming from the Drell-Yan process $\bar{p}p \rightarrow Z'_\mu$.
To obtain the constraints, we assumed that the first two families have the same $Z'$ charges. }
\label{tabla8}
\end{center}
\end{table}

\begin{center}
\begin{table}
\scalebox{0.70}{
\begin{tabular}{c|c|ccccc}\hline
Model&$j$&SM Lepton Embeddings&Universal |&\du & | Lepton Configuration& | LHC-Lower limit~(TeV)
\\\hline
\multirow{1}{*}{A}&-&  $3S_{2}^{\bar\ell+e^{+}}$&\checkmark&$\times$&$3C_2$
&4.87\\\hline
\multirow{1}{*}{B}&-&  $3S_{1}^{\ell+e^{+}}$&\checkmark&$\times$&$3C_1$
&5.53\\\hline
\multirow{2}{*}{C$^j$}&1&  $S_{1}^{\ell+e^{+}}+S_{2}^{\bar\ell+e^{+}}+S_{9}^{\bar\ell+e^{\prime+}}$&$\times$&$\times$&$C_1+C_2+C_3$
&\\
&2&$(S_{1}^{\ell}+S_{9}^{e^{\prime+}})+S_{2}^{\bar\ell+e^{+}}+(S_{9}^{\bar\ell}+S_{1}^{e^{+}})$&$\times$&\checkmark& $2C_2+C_4$
&4.87\\\hline
\multirow{2}{*}{D$^j$}&1&  $S_{1}^{\ell+e^{+}}+S_{2}^{\bar\ell+e^{+}}+S_{10}^{\ell+e^{+}}$&$\times$&\checkmark& $2C_1+C_2$
&5.53\\
&2&$S_{1}^{\ell+e^{+}}+S_{10}^{2\ell+2e^{+}}$&\checkmark&$\times$& $3C_1$
&5.53\\\hline
\multirow{8}{*}{E$^j$}
&1&$2S_{9}^{\bar\ell+e^{\prime +}}+S_{10}^{\ell+e^{+}}$&$\times$&\checkmark& $C_1+2C_3$
&5.75\\
&2&$S_{9}^{2\bar\ell}+S_{10}^{\ell+3e^{+}}$&$\times$&\checkmark& $C_1+2C_2$&4.98\\
&3&  $S_{9}^{2\bar\ell+e^{\prime +}}+S_{9}^{\bar\ell+e^{\prime +}}
+S_{10}^{e^{+}}$&$\times$&\checkmark& $C_2+2C_3$
&5.75\\
&4&$S_{9}^{2\bar\ell+e^{\prime +}}+S_{9}^{\bar\ell}
+S_{10}^{2e^{+}}$&$\times$&\checkmark&$2C_2+C_3$
&4.98\\
&5&$S_{9}^{2\bar\ell}+S_{9}^{\bar\ell}
+S_{10}^{3e^{+}}$&\checkmark&$\times$& $3C_2$
&4.98\\
&6&$S_{9}^{2\bar\ell+e^{\prime +}}+S_{10}^{\ell+2e^{+}}$&$\times$&$\times$& $C_1+C_2+C_3$
&\\
&7&$(S_9^{2\ell+e^{\prime +}}+S_{10}^{e^{+}})+(S_{10}^{\ell}+S_9^{e^{\prime +}})$&$\times$&$\times$&$C_2+C_3+C_4$
&\\
&8&$(S_9^{2\ell}+S_{10}^{2e^{+}})+(S_{10}^{\ell}+S_9^{e^{\prime +}})$&$\times$&\checkmark& $2C_2+C_4$
&4.98\\
&9&$S_{9}^{\bar\ell+e^{\prime +}}+S_{10}^{2\ell+2e^{+}}$&$\times$&\checkmark&$2C_1+C_3$
&5.53\\
&10&$S_9^{\ell+e^{\prime +}}+(S_{10}^{2\ell+e^{+}}+S_9^{e^{\prime +}})$&$\times$&$\times$&$C_1+C_3+C_4$
&\\
&11&$S_{9}^{\bar\ell}+S_{10}^{2\ell+3e^{+}}$&$\times$&\checkmark&$2C_1+C_2$
&5.53\\
&12&$(S_{9}^{\bar\ell}+S_{10}^{e^+})+(S_{10}^{2\ell+e^+}+S_9^{e^{\prime +}})$&$\times$&$\times$& $C_1+C_2+C_4$
&\\
&13&$(S_{9}^{\bar\ell}+S_{10}^{e^+})+(S_{10}^{2\ell}+2S_9^{e^{\prime +}})$&$\times$&\checkmark& $C_2+2C_4$
&7.0\\\hline
\multirow{6}{*}{F$^j$}
&1&  $S_{9}^{\bar\ell+e^{\prime +}}+S_{10}^{2\ell+2e^{+}}$&$\times$&\checkmark& $2C_1+C_3$
&5.53\\
&2& $(S_{9}^{\bar\ell}+S_{10}^{e^{+}})+S_{10}^{2\ell+2e^{+}}$&$\times$&\checkmark&$2C_1+C_2$
&5.53\\
&3& $(S_{9}^{\bar\ell}+S_{10}^{e^{+}})+(S_{10}^{2\ell+e^{+}}+S_{9}^{e^{\prime +}})$&$\times$&$\times$& $C_1+C_2+C_4$
&\\
&4& $(S_{9}^{2\bar\ell+e^{\prime +}}+S_{10}^{e^{+}})+S_{10}^{\ell+e^{+}}$&$\times$&$\times$&$C_1+C_2+C_3$
&\\
&5& $(S_{9}^{2\bar\ell}+S_{10}^{2e^{+}})+S_{10}^{\ell+e^{+}}$&$\times$&\checkmark&$C_1+2C_2$
&4.98\\
&6& $(S_{9}^{2\bar\ell}+S_{10}^{2e^{+}})+(S_{10}^{\ell}+S_{9}^{e^{\prime +}})$&$\times$&\checkmark& $2C_2+C_4$
&4.98\\
&7& $S_{10}^{2\ell+2e^+}+S_{10}^{\ell+e^+}$&\checkmark&$\times$&$3C_1$
&5.53\\
&8& $S_{10}^{2\ell+2e^+}+S_{10}^{\ell}+S_{9}^{e^{\prime +}}$&$\times$&\checkmark&$2C_1+C_4$
&5.53\\\hline
\multirow{1}{*}{G}&-&  $3S_{9}^{\bar\ell+e^{\prime +}}$&\checkmark&$\times$&$3C_3$
&5.53\\\hline
\multirow{1}{*}{H}&-&  $3S_{10}^{\ell+e^{+}}$&\checkmark&$\times$&$3C_1$
&5.53\\\hline
\multirow{3}{*}{I$^j$}&1&  $3S_{2}^{\bar\ell+e^{+}}$&\checkmark&$\times$&$3C_2$
&4.87\\
&2&  $2S_{2}^{\bar\ell+e^{+}}+S_{2}^{\bar\ell}+S_5^{e^{\prime +}}$&$\times$&\checkmark& $2C_2+C_3$
&4.87\\
&3&  $S_{2}^{\bar\ell+e^{+}}+2S_{2}^{\bar\ell}+2S_5^{e^{\prime +}}$&$\times$&\checkmark& $C_2+2C_3$
&5.53\\
&4&  $3S_{2}^{\bar\ell}+3S_5^{e^{\prime +}}$&\checkmark&$\times$&$3C_3$
&5.53\\\hline
\multirow{1}{*}{J}&-&  $3S_{1}^{\ell+e^{+}}$&\checkmark&$\times$&$3C_1$
&5.53\\\hline
\end{tabular}
}
\caption{\footnotesize
Alternative embeddings for the classical AFSs. The superscripts correspond to the particle content of the SM, where $\ell$~($\bar{\ell}$) 
stands for  a left-handed lepton doublet embedded in a $SU(3)_L$ triplet~(anti-triplet), 
and $e^{\prime +}$~($e^+$) is the right-handed charged lepton embedded in a $SU(3)_L$ triplet~(singlet). The lepton content of the model $C_i$ 
was defined in Table~\ref{tabla8}. 
The check mark $\checkmark$ means that at least two families (2+1) or three families (universal) have the same charges under the gauge symmetry, the cross $\times$ stands for the opposite.
LHC constraints are obtained  from Table~\ref{tabla8} for embeddings for which we can choose the same $Z'$ charges for the first two families,
otherwise we leave the space blank.}
\label{tabl5}
\end{table}
\end{center}

\newpage

\begin{center}
\begin{table}
\scalebox{0.70}{
\begin{tabular}{c|c|ccccc}\hline
Model&j&SM Lepton Embeddings&Universal |&\du &| Lepton Configuration
&| LHC-Lower limit~(TeV)\\\hline
\multirow{4}{*}{$Q^{Ij}_3$}&1&  $3S_2^{\bar\ell+e^+}$&\checkmark&$\times$& $3C_2$
&4.87\\
&2&  $2S_2^{\bar\ell+e^+}+(S_2^{\bar\ell}+S_7^{e^{\prime +}})$&$\times$&\checkmark& $2C_2+C_3$
&4.87\\
&3&  $S_2^{\bar\ell+e^+}+(2S_2^{\bar\ell}+2S_7^{e^{\prime +}})$&$\times$&\checkmark& $C_2+2C_3$
&5.53\\
&4&  $3S_2^{\bar\ell}+3S_7^{e^{\prime +}}$&\checkmark&$\times$&$3C_3$
&5.53\\\hline
\multirow{19}{*}{$Q^{Ij}_7$}
&1&  $3S_1^{\ell+e^+}$&\checkmark&$\times$&$3C_1$
&5.33\\
&2&  $2S_1^{\ell+e^+}+(S_1^{\ell}+S_7^{e^{\prime +}})$&$\times$&\checkmark&$2C_1+C_4$
&5.33\\
&3&  $S_1^{\ell+e^+}+(2S_1^{\ell}+2S_7^{e^{\prime +}})$&$\times$&\checkmark& $C_1+2C_4$
&6.52\\
&4&  $3S_1^{\ell}+3S_7^{e^{\prime +}}$&\checkmark&$\times$&$3C_4$
&6.52\\
&5&$2S_1^{\ell+e^+}+S_7^{\bar\ell+e^{\prime +}}$&$\times$&\checkmark&$2C_1+C_3$
&5.33\\
&6&$2S_1^{\ell+e^+}+(S_7^{\bar\ell}+S_1^{e^+})$&$\times$&\checkmark&$2C_1+C_2$
&5.33\\
&7&$S_1^{\ell+e^+}+(S_1^{\ell}+S_7^{e^{\prime +}})+S_7^{\bar\ell+e^{\prime +}}$&$\times$&$\times$&$C_1+C_3+C_4$
&\\
&8&$S_1^{\ell+e^+}+(S_1^{\ell}+S_7^{e^{\prime +}})+(S_7^{\bar\ell}+S_1^{e^+})$&$\times$&$\times$&$C_1+C_2+C_4$
&\\
&9&$(2S_1^{\ell}+2S_7^{e^{\prime +}})+S_7^{\bar\ell+e^{\prime +}}$&$\times$&\checkmark& $C_3+2C_4$
&6.52\\
&10&$(2S_1^{\ell}+2S_7^{e^{\prime +}})+(S_7^{\bar\ell}+S_1^{e^+})$&$\times$&\checkmark&$C_2+2C_4$
&6.52\\
&11&$S_1^{\ell+e^+}+2S_7^{\bar\ell+e^{\prime +}}$&$\times$&\checkmark&$C_1+2C_3$
&5.53\\
&12&$S_1^{\ell+e^+}+S_7^{\bar\ell+e^{\prime +}}+(S_7^{\bar\ell}+S_1^{e^+})$&$\times$&$\times$& $C_1+C_2+C_3$
&\\
&13&$S_1^{\ell+e^+}+(2S_7^{\bar\ell}+2S_1^{e^+})$&$\times$&\checkmark&$C_1+2C_2$
&4.87\\
&14&$(S_1^{\ell}+S_7^{e^{\prime +}})+2S_7^{\bar\ell+e^{\prime +}}$&$\times$&\checkmark& $2C_3+C_4$
&5.53\\
&15&$(S_1^{\ell}+S_7^{e^{\prime +}})+S_7^{\bar\ell+e^{\prime +}}+(S_7^{\bar\ell}+S_1^{e^+})$&$\times$&$\times$& $C_2+C_3+C_4$
&\\
&16&$(S_1^{\ell}+S_7^{e^{\prime +}})+(2S_7^{\bar\ell}+2S_1^{e^+})$&$\times$&\checkmark&$2C_2+C_4$
&4.87\\
&17&$3S_7^{\bar\ell+e^{\prime +}}$&\checkmark&$\times$&$3C_3$
&5.53\\
&18&$2S_7^{\bar\ell+e^{\prime +}}+(S_7^{\bar\ell}+S_1^{e^+})$&$\times$&\checkmark&$C_2+2C_3$
&5.53\\
&19&$S_7^{\bar\ell+e^{\prime +}}+(2S_7^{\bar\ell}+2S_1^{e^+})$&$\times$&\checkmark&$2C_2+C_3$
&4.87\\
&20&$3S_7^{\bar\ell}+3S_1^{e^+}$&\checkmark&$\times$& $3C_2$
&4.87\\\hline
\end{tabular}}
\caption{Alternative embeddings for new anomaly-free sets. The superscripts correspond to the particle content of the SM, where $\ell$~($\bar{\ell}$) stands for  a left-handed lepton doublet embedded in a $SU(3)_L$ triplet~(anti-triplet), 
and $e^{\prime +}$~($e^+$) is the right-handed charged lepton embedded in a $SU(3)_L$ triplet~(singlet). The lepton content of the model $C_i$ 
was defined in Table~\ref{tabla8}. The check mark $\checkmark$ means that at least two families (2+1) or three families (universal) have the same charges under the gauge symmetry, the cross $\times$ stands for the opposite. LHC constraints are obtained from Table~\ref{tabla8}  for embeddings for which we can choose the same $Z'$ charges for the first two families, otherwise we leave the space blank. 
}
\label{tabl6}
\end{table}
\end{center}
%
\FloatBarrier

\section{Conclusions}
Restricting ourselves to models without exotic electric charges, we have built 12 sets of particles $ S_i $ from triplets, antitriplets and singlets of $ SU(3)_L\otimes U(1)_X$. These sets are constructed in such a way that they contain the charged particles and their respective antiparticles,
following a similar procedure  to that in references~\cite{Ponce:2001jn,Ponce:2002sg}.
With these sets, we built the IAFSs~$ L_i $, $Q_i^{I}$, $Q_i^{II}$ and $Q_i^{III}$ depending on their quark content, as it is shown in Table III. From the IAFSs it is possible to systematically build  3-3-1 models. It is important to realize that if we restrict the AFSs to a minimum content of vector-like structures (i.e, $L_i$), having a lepton and quark sector consistent with the SM,  our analysis is reduced to the AFSs that containt the  classical 3-3-1 models reported in~\cite{Ponce:2001jn,Ponce:2002sg}. However, if we allow alternative embeddings for SM particles within $S_i$, we get new phenomenological distinguishable  models.
Table~\ref{tabl5} lists all the possible embeddings for the sets of fermions that originate the models reported in references~\cite{Ponce:2001jn,Ponce:2002sg}. In Tables~\ref{tabl5} and \ref{tabl6},  $C^1$,  $D^1$,  $E^1 $,  $F^1 $ and $I^1$ correspond to the models {\bf C}, {\bf D}, {\bf E}, {\bf F } and {\bf I} in references~\cite{Ponce:2001jn,Ponce:2002sg}. 

By combining the IAFSs from Table~\ref{tabl3}, it is possible to find a large number of models. 
By restricting to models with a minimal content of exotic fermions, we found 1682 models which could be of phenomenological interest. 
To exemplify the new realistic AFSs that can be formed, we reported  some of them, with their corresponding embeddings, in Table~\ref{tabl6}.
We also report LHC constraints  for models with the first two families having SM fermions with identical charges, including some of the classical 3-3-1 models,  as reported in Table~\ref{tabla8}. From this Table we can see that, independent of the model, the mass value of the new neutral gauge boson for all the 3-3-1 models considered in this paper, is above 4.87 TeV.
\FloatBarrier

\section{Acknowledgments}
R. H. B. and L. M. thank the ``Dirección de investigaciones ITM" project number P20246. We thank Financial support from “Patrimonio
Autónomo Fondo Nacional de Financiamiento para la Ciencia, la Tecnología y la Innovación, Francisco José de Caldas”, and “Sostenibilidad-UDEA”. This research was partly supported by the  ``Vicerrector\'ia de Investigaciones e Interacci\'on Social VIIS de la Universidad de Nari\~no'', project numbers 1928 and 2172.

\appendix 
\section{\( Z'\) Charges}
\label{sec:apendice1}

\begin{table}[h!]
\begin{center}
\begin{tabular}{| c | c | c |}
\hline
\multicolumn{3}{|c|}{$C_1$,\ \ \  $\ell\subset 3$, $e\subset 1$  (as in $S_1$) }\\
\hline
Fields & Vectorial & Axial \\ \hline
$\nu_e$ & $ -\frac{1}{2} \left(\frac{\cos \theta}{\delta}  +\sin\theta\right)$ & $ -\frac{1}{2} \left(\frac{\cos \theta}{\delta}  +\sin\theta\right)$  \\ \hline
$e$ & $ \frac{1}{2} \left[\sin \theta  (1-4\sin^2\theta_W)-\frac{\cos \theta}{\delta} (1+2\sin^2\theta_W)\right]$ & $ \frac{1}{2} \left[\sin \theta -\frac{\cos \theta}{\delta} (1-2\sin^2\theta_W)\right]$ \\ \hline
\hline
\end{tabular}
\caption{$\bar{f}f \rightarrow Z'_\mu$}
\label{tab:c1}
\end{center}
\end{table}

\begin{table}[h!]
\begin{center}
\begin{tabular}{| c | c | c |}
\hline
\multicolumn{3}{|c|}{$C_2$,\ \ \  $\ell\subset  3^*$, $e\subset 1$ (as in $S_2$)  }\\ 
\hline
Fields & Vectorial & Axial \\ \hline
$\nu_e$ & $\frac{\cos\theta}{\delta}\left(\frac{1}{2}-\sin^2\theta_W\right) - \frac{\sin\theta}{2}$ & $\frac{\cos\theta}{\delta}\left(\frac{1}{2}-\sin^2\theta_W\right) - \frac{\sin\theta}{2}$ \\ \hline
$e$ & $\sin\theta\left(\frac{1}{2}-2\sin\theta_W^2\right)+\frac{\cos\theta}{\delta}\left(\frac{1}{2}-2\sin\theta_W^2\right)$ & $\frac{1}{2}\left(\frac{\cos\theta}{\delta}+\sin\theta\right)$ \\ \hline
 \hline
\end{tabular}
\caption{$\bar{f}f \rightarrow Z'_\mu$}
\label{tab:c2}
\end{center} 
\end{table}

\begin{table}[h!]
\begin{center}
\begin{tabular}{| c | c | c |}
\hline
\multicolumn{3}{|c|}{$C_3$, \ \ \ $\ell\subset 3^*$, $e\subset  3^*$ (as in $S_7$) }\\
\hline
Fields & Vectorial & Axial \\ \hline
$\nu_e$ & $ \frac{1}{2} \left[\frac{\cos \theta} {\delta} (1-2\sin^2 \theta_W) -\sin\theta\right] $ & $ \frac{1}{2} \left[\frac{\cos \theta} {\delta} (1-2\sin^2 \theta_W) -\sin\theta\right] $ \\ \hline
$e$ & $ \sin\theta\left(\frac{1}{2}-2\sin^2\theta_W\right)+\frac{3\cos\theta}{2\delta}(1-2\sin^2\theta_W) $& $ \frac{1}{2}\left[ \sin\theta-\frac{\cos\theta}{\delta}(1-2\sin^2\theta_W) \right]$ \\ \hline
\hline
\end{tabular}
\caption{$\bar{f}f \rightarrow Z'_\mu$}
\label{tab:c3}
\end{center}
\end{table}

\begin{table}[h!]
\begin{center}
\begin{tabular}{| c | c | c |}
\hline
\multicolumn{3}{|c|}{$C_4$,\ \ \  $\ell\subset 3$, $e\subset 3^*$}\\
\hline
Fields & Vectorial & Axial \\ \hline
$\nu_e$ 
& $\frac{1}{2}\left(\sin\theta+\frac{\cos\theta}{\delta}\right) $ 
& $\frac{1}{2}\left(\sin\theta+\frac{\cos\theta}{\delta}\right) $  \\ \hline
$e$ 
& $ \frac{1}{2}\left[\sin\theta(1-4\sin^2\theta_W)+\frac{\cos\theta}{\delta}(1-2\sin^2\theta_W)\right]$ 
& $ \frac{1}{2}\left[\sin\theta(1-2\sin^2\theta_W)-\frac{3\cos\theta}{\delta}\right]$ \\ \hline \hline
\end{tabular}
\caption{$\bar{f}f \rightarrow Z'_\mu$}
\label{tab:c4}
\end{center}
\end{table}

\begin{table}[h!]
\begin{center}
\begin{tabular}{| c | c | c |}
\hline
\multicolumn{3}{|c|}{$\bar{q}$,\ \ \ $q\subset 3^*$ (as in $S_3$) }\\
\hline
Fields & Vectorial & Axial \\ \hline
$u$ & $ \frac{1}{6}\left[\sin\theta \left(-3+8\sin^2\theta_W\right)+\frac{\cos\theta}{\delta}\left(3+2\sin^2\theta_W\right)\right] $ & $\frac{1}{2}\left[\frac{\cos\theta}{\delta}\left(1-2\sin^2\theta_W\right)-\sin\theta\right]$ \\ \hline
$d$ & $\left[\frac{1}{2}-\frac{2}{3}\sin^2\theta_W\right]\left[\sin\theta+\frac{\cos\theta}{\delta}\right] $ & $ \frac{1}{2}\left[\sin\theta+\frac{\cos\theta}{\delta}\right]$ \\ \hline
\hline
\end{tabular}
\caption{$\bar{f}f \rightarrow Z'_\mu$}
\label{tab:qbar}
\end{center}
\end{table}

\begin{table}[h!]
\begin{center}
\begin{tabular}{| c | c | c |}
\hline
\multicolumn{3}{|c|}{ $q$,\ \ \ $q\subset 3$ (as in $S_4$) }\\
\hline
Fields & Vectorial & Axial \\ \hline
$u$ & $ \frac{1}{6}\left[\sin\theta \left(-3+8\sin^2\theta_W\right)+\frac{\cos\theta}{\delta}\left(5-8\cos^2\theta_W\right)\right] $ & $-\frac{1}{2}\left[\frac{\cos\theta}{\delta}+\sin\theta\right]$ \\ \hline
$d$ & $ \frac{1}{6}\left[\sin\theta \left(3-4\sin^2\theta_W\right)-\frac{\cos\theta}{\delta}\left(3-2\sin^2\theta_W\right)\right] $ & $ \frac{1}{2}\left[\sin\theta-\frac{\cos\theta}{\delta}\left(1-2\sin^2\theta_W\right)\right]$ \\ \hline
\hline
\end{tabular}
\caption{$\bar{f}f \rightarrow Z'_\mu$}
\label{tab:q}
\end{center}
\end{table}

where $\theta$ is an angle mixing between $Z$ and $Z'$ bosons, $\theta_W$ is the Weinberg angle and $\delta=\sqrt{4\cos^2\theta_W-1}$.

\FloatBarrier

\end{document}